\documentclass[a4paper]{jpconf}
\usepackage{graphicx}
\begin{document}
\begin{flushright}
UCLA/05/TEP/33\\
December 2005
\end{flushright}

\vspace{-1.5truecm}

\title{DAMA detection claim  is still compatible with all other DM
        searches\footnote{Talk given at TAUP2005, Sept. 10-14 2005, Zaragoza (Spain)}}

\author{Graciela B Gelmini}

\address{Department of Physics and Astronomy, UCLA, Los Angeles,
CA 90095-1547, USA}

\ead{gelmini@physics.ucla.edu}

\begin{abstract}
We show that the annual modulation signal observed by DAMA can be
reconciled with all other negative results from dark
matter searches with a conventional halo model
for particle masses around 5 to 9 GeV. We also show which particular
dark matter stream  could produce the DAMA signal.
\end{abstract}

Given the country which hosts this conference, TAUP2005, I was tempted to
 change the title into a well known
spanish aphorism ``Los muertos que vos matais gozan the buena salud". Although the DAMA signal may not   be entirely  in``good health", the point of this talk, based on  Ref.~\cite{GG},  is to show that, even with the usual 
assumptions on the dark halo of our galaxy, the DAMA claim has not yet been rejected by other experimental results. 

The DAMA collaboration \cite{DAMA}, has found an
annual modulation in its data compatible with the signal expected from
dark matter (DM) particles bound to our galactic halo. Other such experiments,  CDMS \cite{CDMS-I,CDMS-II}, EDELWEISS
\cite{EDELWEISS,EDELWEISSfinal}, 
and CRESST \cite{CRESST-I,CRESST-II}, have not found
any signal from weakly interacting massive particles (WIMPs). It has been difficult to reconcile a WIMP
signal in DAMA with the other negative results~\cite{previous}.

In  Ref.~\cite{GG} we showed that it is possible to have a dark matter signal above
the WIMP speed threshold for DAMA and below the WIMP speed threshold
for CDMS and EDELWEISS, so that the positive and negative detection
results can be compatible.  Our idea is based on attributing the
  DAMA signal to scattering
  off Na, instead of I, and can be tested in the immediate future by
  detectors using light nuclei, such as CDMS-II (using Si) and
  CRESST-II (using O).
 We found: (1) that with the standard dark
halo model there is a solution for WIMP masses about 5-9 GeV and
WIMP-proton scattering cross section of about 1 femtobarn
($10^{-39}$~cm$^2$), and (2) that this region of solutions can be
enlarged if a dark matter stream is suitably added to the standard
dark halo. DM streams have been considered before~\cite{streams},
 but we looked for a stream with
speed above threshold for Na in DAMA and below threshold for Ge in
CDMS and with the right arrival direction to produce the annual modulation seen by DAMA. 
The stream should come from the general direction of the Galactic rotation
  (it cannot be the Sagittarius stream). See Figs. 1 and 2.  The region in point (1) could certainly also be enlarged by
considering more general halo models, even in the absence of dark
matter streams (see e.g.\ the models in~\cite{DAMA03}).

Light neutralinos as WIMPs with masses as low as 2~GeV
\cite{stodolsky} or, with updated bounds, 6~GeV \cite{bottino} have
been considered, but their cross sections are about one order of
magnitude smaller than those needed here. In Ref~\cite{GG} we proceeded in
a purely phenomenological way in choosing the WIMP mass and cross
section. For simplicity we concentrated on spin-independent cross sections only.

Our basic idea is very simple. The minimum WIMP speed required to produce a  nuclear
recoil energy  $E$ is $v=\sqrt{\frac{ME}{2\mu^2}}$ where $\mu = {\frac{m M}{(m+M)}}$ is the reduced
mass,  $m$ the WIMP mass and  $M$ the nucleus mass. Thus the experimental energy threshold
$E_{\rm th}$  corresponds to a minimum observable speed  $v_{\rm th}$. Let us simplify our argument
by considering $m \ll M$, then $\mu \simeq  m$ and $v_{\rm th} \sim \sqrt{ME_{\rm th}}$.
One can easily compare the square-roots for Ge in CDMS-SUF ($\sqrt{67.64~{\rm GeV} \times 5 {\rm keV}}$) and for Na in DAMA ($\sqrt{21.41~{\rm GeV} \times 6.7 {\rm keV}}$), and find that $v_{\rm th-CDMS-SUF-Ge} = 2.44 ~v_{\rm th-DAMA-Na}$. The $v_{\rm th}$ are even  larger 
for CDMS-Soudan, EDELWEISS and
experiments with heavier nuclei. Without using $m \ll M$, one can easily see that $v_{\rm th-CDMS-SUF-Ge} > v_{\rm th-DAMA-Na}$ 
for $m < 22.3$ GeV.
Thus, it  is possible to have halo DM WIMPs with speeds above threshold
for Na in DAMA and below threshold for Ge in CDMS and EDELWEISS.

\begin{figure}
\includegraphics[width=0.45\textwidth]{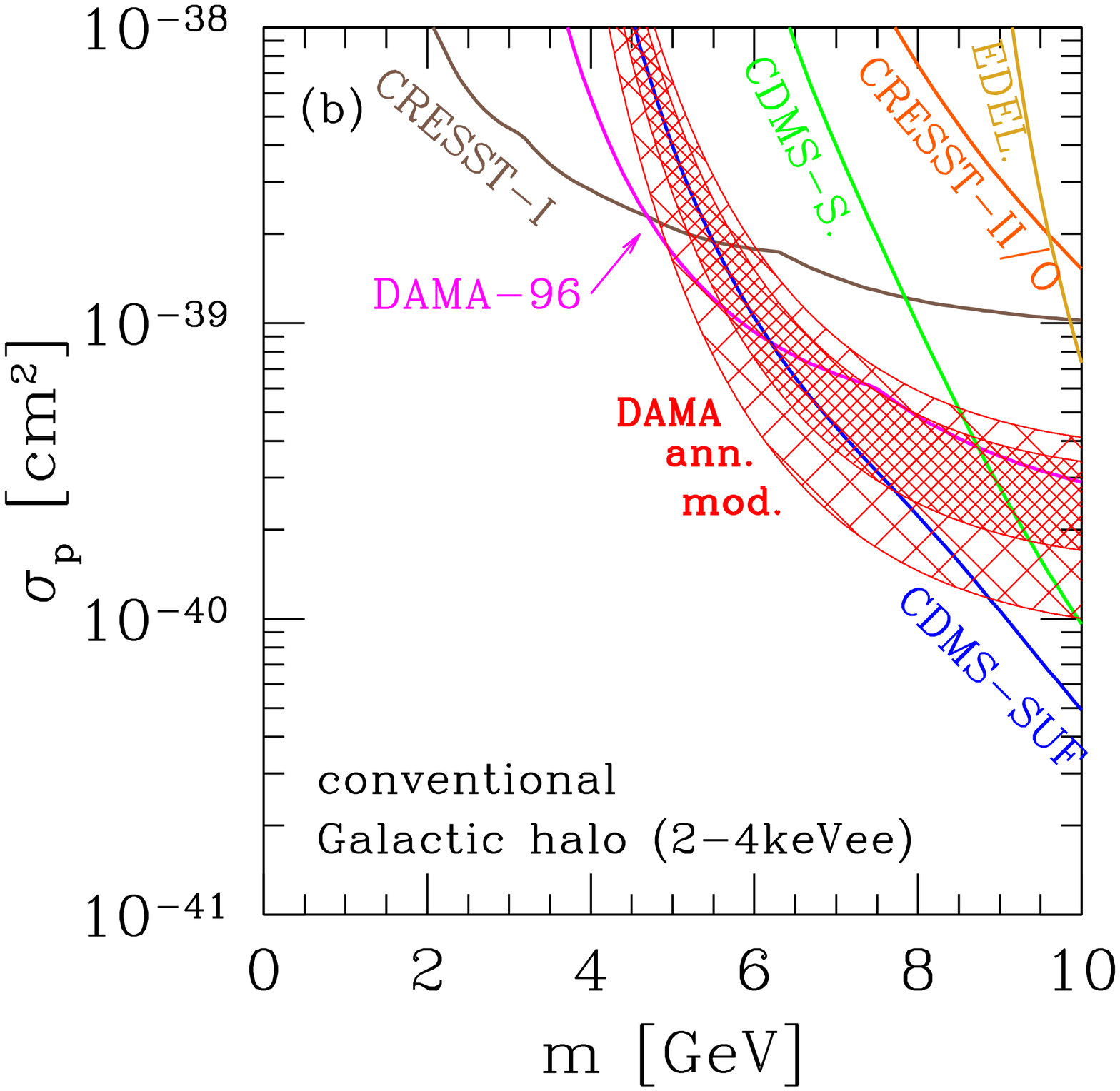}
\includegraphics[width=0.45\textwidth]{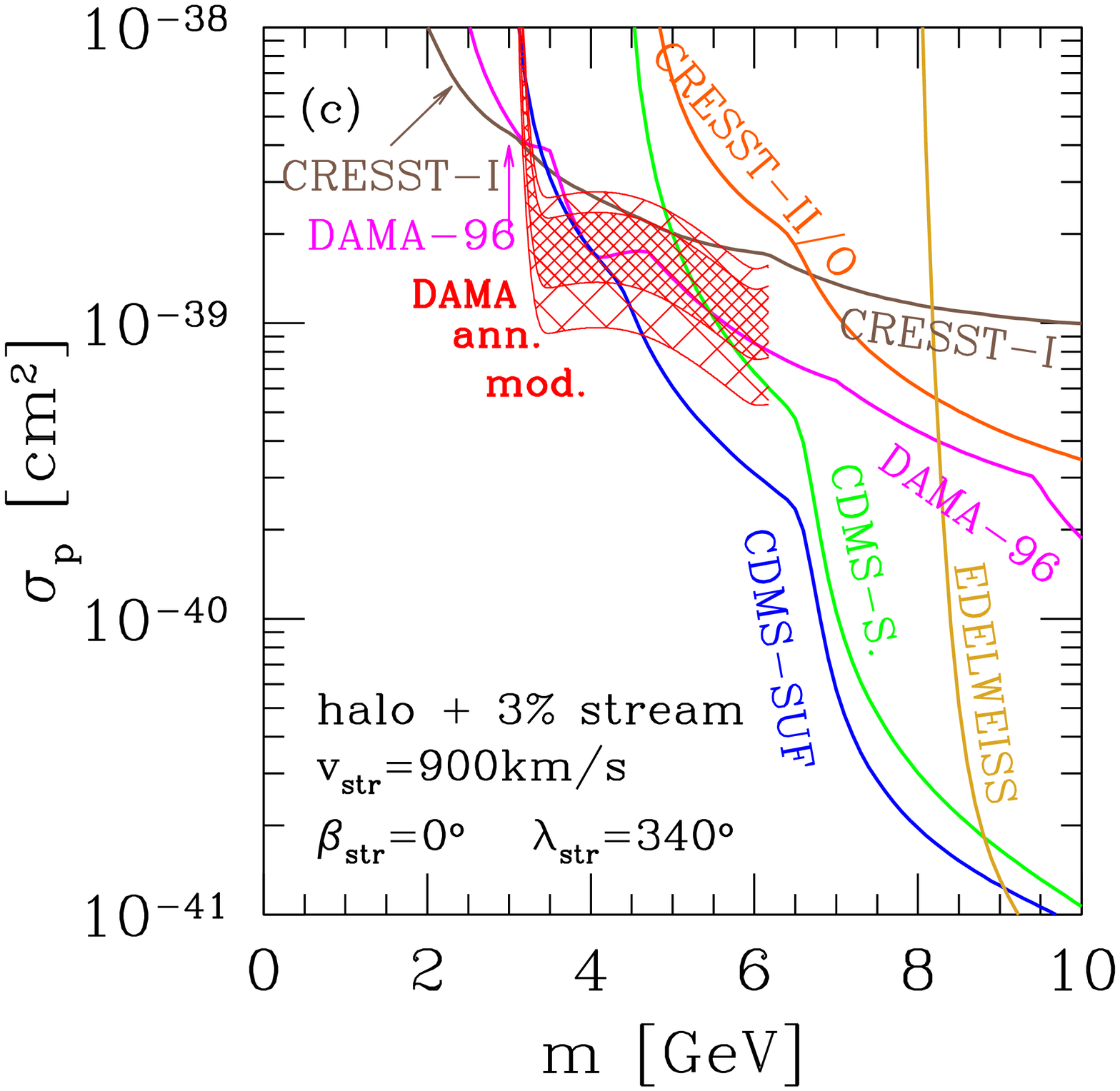}
\caption{Comparison of
  the DAMA annual modulation region with other direct
  detection bounds for spin-independent WIMP-proton interactions (a) (left panel) for just the
  conventional dark halo model, and  (b) (right panel) this halo model with the addition of a DM stream
  with density 3\% of the conventional local halo density,
  heliocentric arrival direction of ecliptic coordinates
  $(\lambda_{\rm str}, \beta_{\rm str})=(340^\circ, 0^\circ)$, and
  heliocentric speed of  900~km/s. In the hatched
  region, the WIMP-proton cross section $\sigma_{\rm p}$ at WIMP mass
  $m$ reproduces the DAMA annual modulation results at the 90\% and
  3$\sigma$ C.L. (inner densely hatched region and outer hatched
  region, respectively).  The region above each other line is excluded
  at 90\% C.L. by the corresponding experiment (CDMS-Soudan is denoted by CDMS-S).
   In (a), we use the 2-4 and 6-14 keVee DAMA bins; 
  in (b), the 2-6 and 6-14 keVee bins. 
  In (b) the gap in the DAMA modulation region  is due to our requirement that $\chi^2_{\rm
    min}<2$  (see Ref.~\cite{GG}). The experimental upper limits change when the stream is
  included. }
\end{figure}

\begin{figure}
\includegraphics[width=0.45\textwidth]{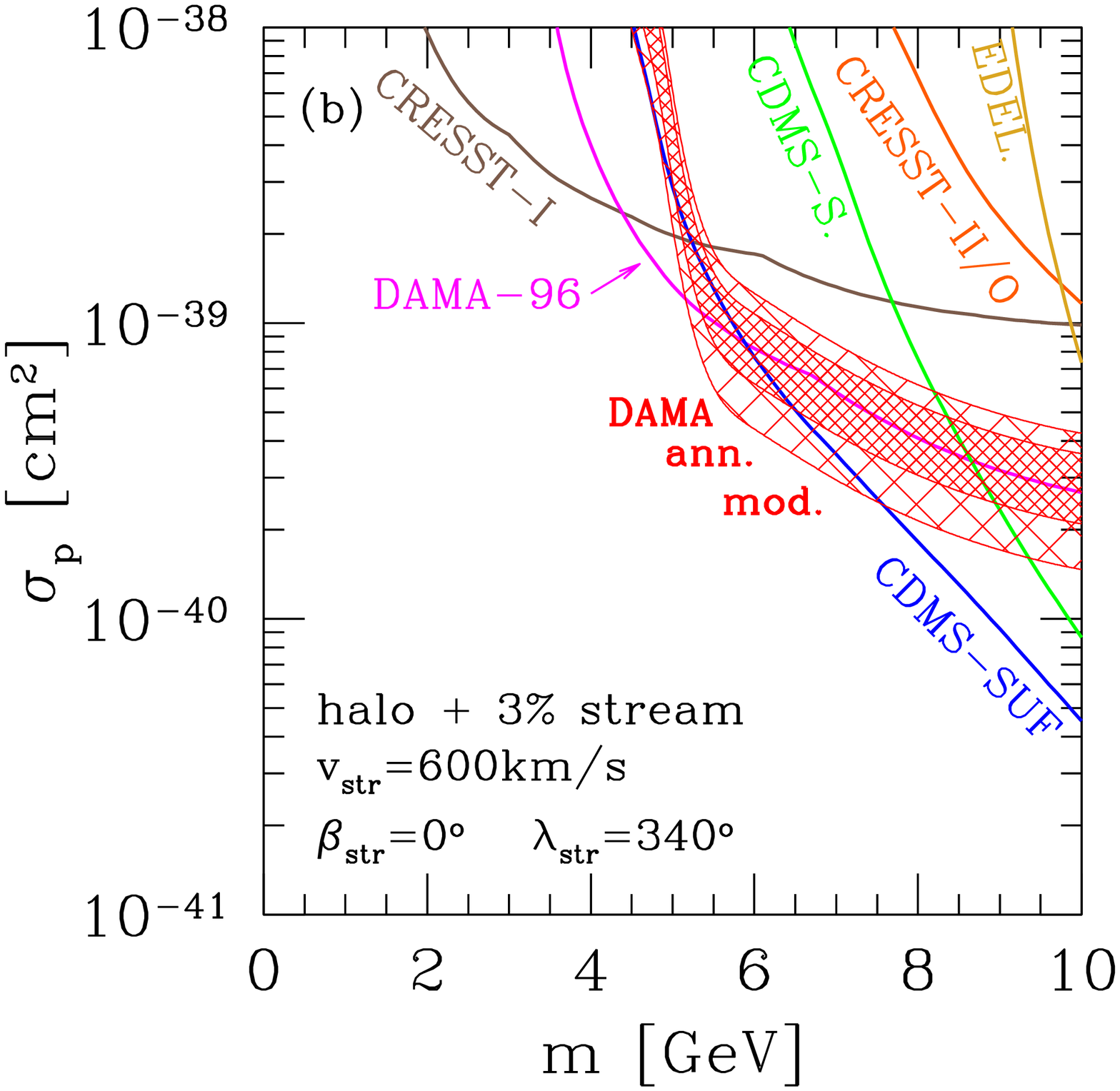}
 \includegraphics[width=0.45\textwidth]{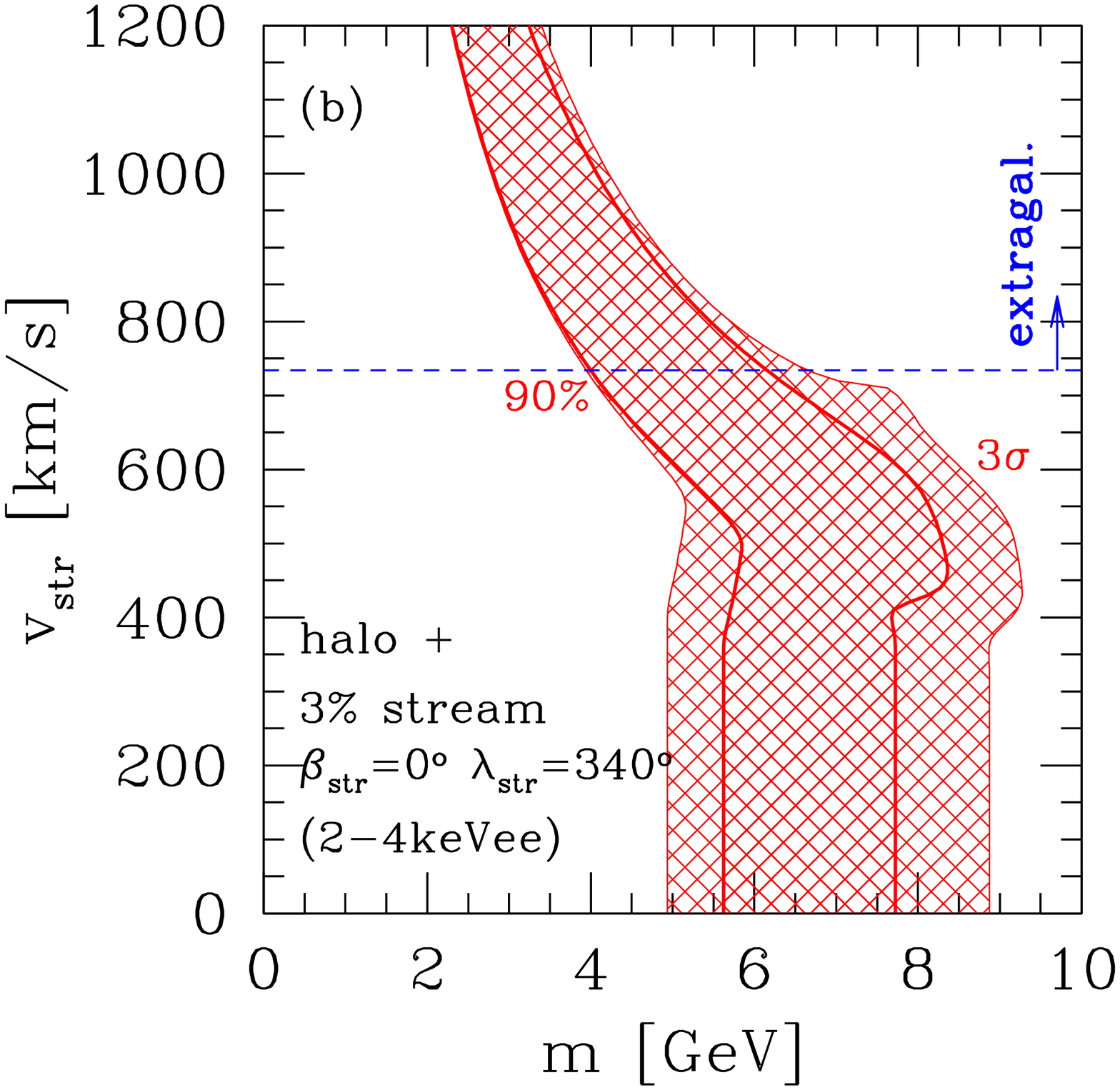}
\caption{(a) (left panel) As in Fig.~1b but with heliocentric stream speed  $v_{\rm str}$ of  600~km/s. (b) (right panel) Range of WIMP masses $m$ for which there is a 
  compatible region between the DAMA modulation and the other
  experimental results at various $v_{\rm
    str}$.   Also indicated is the speed above which
  the stream is extragalactic (dashed horizontal line).  The inner
  densely hatched and outer hatched regions correspond to the 90\% and
  3$\sigma$ C.L., respectively.  
  In (b) the  2-4 and 6-14 keVee DAMA bins were used. 
  At the 3$\sigma$
  level it is possible to find WIMP masses compatible with DAMA and
  all other experiments at any assumed stream speed (see Ref.~\cite{GG}). }
  \end{figure}

Three light nuclei, namely Si in CDMS and Al and O in
CRESST, have speed thresholds lower than Na in DAMA, and can be used to test and constrain our idea. In a recent paper~\cite{Akerib:2005kh}, announced at this conference, CDMS obtained bounds on
light WIMPs using their Si component, which get into part of the
mass region we favor.  

\section*{Acknowledgments}

This work  was supported in part by NASA grant NAG5-13399.
G.G was supported in part by the US DOE grant DE-FG03-91ER40662 
Task C.

\end{document}